# Momentum resolved ground/excited states and the ultra-fast excited state dynamics of monolayer MoS$_2$

Woojoo Lee[1], Yi Lin[2], Li-Shuan Lu[3], Wei-Chen Chueh[3], Mengke Liu[1], Xiaoqin Li[1], Wen-Hao Chang[3,4], Robert A. Kaindl[2,5], and Chih-Kang Shih[1,†]

[1] Department of Physics, The University of Texas at Austin, Austin, Texas 78712, USA

[2] Materials Sciences Division, Lawrence Berkeley National Laboratory, Berkeley, California 94720, USA

[3] Department of Electrophysics, National Chiao Tung University, Hsinchu 30010, Taiwan

[4] Center for Emergent Functional Matter Science (CEFMS), National Chiao Tung University, Hsinchu 30010, Taiwan

[5] Department of Physics and CXFEL Laboratory, Arizona State University, Tempe, Arizona 85287, USA

[†] *Corresponding author: shih@physics.utexas.edu*

The emergence of transition metal dichalcogenides (TMD) as crystalline atomically thin semiconductors has created a tremendous amount of scientific and technological interest. Many novel device concepts have been proposed and realized (*1–3*). Nonetheless, progress in k-space investigations of ground/excited state electronic structures has been slow due to the challenge to create large scale, laterally homogeneous samples. Taking advantage of recent advancements in chemical vapor deposition, here we create a wafer-size MoS$_2$ monolayer with well-aligned lateral orientation for advanced electron spectroscopy studies (*4–6*). Low energy electron diffraction and scanning tunneling microscopy (STM) demonstrate atomically clean surfaces with in-plane crystalline orientation. The ground state and excited state electronic structures are probed using scanning tunneling spectroscopy (STS), angle-resolved photoemission (ARPES) and time-resolved (tr-)ARPES. In addition to mapping out the momentum-space quasiparticle band structure in the

**valence and conduction bands, we unveil ultrafast excited state dynamics, including inter- and intra-valley carrier scattering and a rapid downward energy shift by ~ 0.2eV lower than the initial free carrier state at Σ point.**

**Introduction**

Monolayers of semiconducting transition-metal dichalcogenides (TMDs) and their heterostructures have emerged as a powerful system to design novel devices for scientific and technological applications. Although many novel device concepts have been realized in recent years (*1–3*), realizing a scalable and high-quality materials platform remains an outstanding challenge. Scientifically, the difficulty in preparing large area, well-oriented monolayer TMDs has slowed the progress in utilizing momentum space probe of electronic structures except in some limited cases, for example, molecular beam epitaxial growth of $MoSe_2$ and $MoS_2$ on bilayer graphene (*7, 8*). Recently, progress has been made in creating large area monolayer TMDs by exfoliation thus enabling applications of advanced electronic structure probing tools (*9, 10*). Nevertheless, the tedious exfoliation step remains a bottleneck and limits its wide applicability.

Recently, there has been significant progress made in creating wafer scale, highly oriented TMD monolayers on a sapphire substrate using chemical vapor deposition (CVD) (*4–6*), albeit utilization of such platform for advanced electron spectroscopy has yet to be demonstrated. Taking advantage of these recent developments of CVD growth, we demonstrate the ability to create a large scale, atomically clean monolayer TMD (e.g. $MoS_2$) on graphite substrate for electron spectroscopy investigations. More importantly, using a collection of electron microscopy and spectroscopy tools, we provide a consistent picture of the ground state electronic structures and access the otherwise unoccupied excited state dynamics. Scanning tunneling microscopy (STM) and low energy electron diffraction (LEED) reveal a wafer-scale, well-aligned crystallographic orientation. The high-quality sample enables us to apply angle-resolved photoemission (ARPES) and time-resolved (tr-)ARPES to map out ground/excited state electronic structures with high energy and momentum resolution. The capability of observing both the momentum resolved valence band and conduction band states in tr-ARPES permits the direct determination of the quasi-particle band gap. Remarkably, all the critical points energy locations are corroborated with the scanning tunneling spectroscopy. Moreover, we access the dynamical behavior of excited states, including inter- and intra-valley carrier scattering and a rapid energy shift (~0.2 eV) of quasiparticles at the $\Sigma$ point.

**Results**

Figure 1 depicts the process in preparing the large scale well-aligned monolayer $MoS_2$ for advanced electronic structure probe. First, we grow monolayer TMDs on a 2" sapphire wafer using CVD. The growth is based on a "substrate guided growth" procedure reported previously (*4–6*) but with significant refinement to reduce defect density and obtain monolayer TMDs with well-aligned orientation, i.e., all domains have either the same orientation or its mirror twin (Fig. S1). Figure 1a shows a slice (5 cm x 1 cm) of the wafer after monolayer $MoS_2$ growth. Depending on the location at the gas flow stream, the growth leads to either a continuous film (region A), a densely covered region (region B showing 80 ~ 90% coverage), or a low coverage (region C).

For STM, ARPES and tr-ARPES studies, the sample needs to have an electrical contact to complete the circuit loop when the electron is injected or ejected. While previous experiments mostly used side electrical contacts on a continuous film, the presence of a Schottky barrier at the contacts often induces a lateral field (*11*, *12*) along with the carrier motion. Moreover, upon excitation of photoelectrons, transient field can build up (i.e. a photovoltage effect), introducing complications in the interpretation of experimental observations (*9*, *13*, *14*). Finally, in the STM investigation, there will be significant tip-induced band bending along the lateral direction (*15*), hindering a comparison between different electron spectroscopic studies. To circumvent these limitations, we have chosen to transfer the sample to a highly oriented pyrolytic graphite (HOPG) or a graphene substrate which prevents the formation of a surface photovoltage during photoemission and eliminates the tip-induced band bending during STM operation (*16, 17*).

Region B in Figure 1a is chosen (instead of the continuous film region A) so that the contrast in optical images can be used to gauge the success in sample transfer. Detailed description of sample transfer can be found in the Supplementary (Fig. S2). We obtain a large sample size (0.4" x 0.4") limited by the sample holder, which is essential for subsequent measurements. The LEED image shows a well-defined pattern of six dots, confirming the high crystalline quality of the $MoS_2$ monolayers including their twin pairs. The ring pattern is produced by the HOPG. The presence of the twin pairs prevents the spin or degenerate valleys to be resolved in spectroscopy measurements. The same LEED pattern is obtained when the electron beam scans across the sample, indicating a large size, atomically clean surface. The STM measurement was carried out at 77 K, revealing a lattice constant of $3.17 \pm 0.03$ Å, close to the well-accepted value of 3.16 Å. Additionally, ambient stability of $MoS_2$ monolayers on HOPG substrate was also investigated to ensure the consistency of the sample quality when taken out from different chambers to air for different measurements (Fig. S3).

The k-space electronic structures in the valence band are probed using regular ARPES with helium lamp excitation sources of 21.2 and 40.8 eV, as shown in the ARPES maps in Figs. 2a,b and corresponding K point energy distribution curves (EDCs) in Fig. 2c. Each source offers distinct advantages and disadvantages. The 21.2 eV source is brighter but has different sensitivities to dissimilar bands, whereas the 40.8 eV source is weaker (by about two orders of magnitude) but enables a more uniform photoionization cross section for all bands (*18*). Figs. 2d,e show the respective second-derivative images which reveal better contrast for different bands. The states near the VBM at $\Gamma$ and K are well-resolved, revealing the K-valley spin-orbit splitting of $140 \pm 4$ meV and the $\Gamma$ valley located at ~130 meV below the VBM. The deeper valence bands are better revealed by using $h\nu = 40.8\ eV$ where the detailed band structures compare very well with the band structure for a free-standing single layer $MoS_2$ (Fig. 2f) calculated using density functional theory (DFT).

The excited state electronic structure and dynamics are investigated using high-repetition rate, extreme-ultraviolet(XUV) tr-ARPES (*19*). Here we focus on two aspects: (a) direct determination of quasi-particle band gaps based on the excitation into otherwise unoccupied levels, and (b) excited state dynamics at different critical points. The $MoS_2$ monolayer is excited with 2.2-eV pump pulses from a frequency-doubled optical parametric amplifier at 25 kHz repetition rate, with an incident fluence of $\approx 40$ µJ/cm$^2$. In turn, p-polarized, 22.3-eV XUV probe pulses are used for photoemission. Shown in Fig. 3a is the band structure acquired at 300 K at a pump-probe delay time of t= 0.4 ps, where the valence band states and the pump-excited states near the conduction band minimum (CBM) can both be observed. The EDCs taken at the K point at t = 1.27 ps are shown in Fig. 3b with two different sample temperatures (80 K and 300 K). The photoexcitation instantaneously shifts the K-point valence band by ~0.1 eV downward versus the equilibrium state (Fig. S4), then remains relatively steady during the measurement. This effect is observed at both temperatures and its origin will be discussed elsewhere. Since both valence/conduction band states are detected simultaneously, the determination of quasi-particle band gap is not influenced by this effect. At 80 K, the valence band spin-orbit splitting (~ 140 meV) at the K point is well-resolved, but slightly smeared out at 300 K. The CBM at the K point appears at 0.25 eV at 80 K and becomes 0.18 eV at 300 K. The excited-state VBM at K does not change its energy location as a function of temperature beyond the energy resolution (-1.85 eV), yielding a quasi-particle band gap of 2.10 $\pm$ 0.01eV at 80 K and 2.03 $\pm$ 0.01eV at 300 K as shown in Fig. 3b. Shown as the inset is the measured quasi-particle gap value at 80 K as a function of the excitation fluence which shows a very weak trend (if any) of smaller gap at higher fluence.

The determination of the quasi-particle bandgap has been a central issue in TMD for several years (*20*). Many groups have tackled this issue using different techniques but rather inconsistent results (*16, 21–25*). The ability to view simultaneously the valence and conduction band states in the momentum-space using tr-ARPES enables the unambiguous determination. As well, in this study, the acquired quasi-particle band structure using tr-ARPES at 80 K shows excellent agreement with the STS acquired at 77 K on $MoS_2$ on graphite (shown in Fig. 3c) where the VBM is identified at -1.84 eV and CBM at 0.31 eV: with a band gap of 2.15 ± 0.06eV. STS also identifies another threshold at 0.2 ± 0.05 eV above at CBM which is assigned as a local minimum at Σ point (also referred to as Q point (23)). The Σ threshold is confirmed by using a $\frac{\partial Z}{\partial V}$ spectrum acquired at constant current (green curve in Fig. 3c); this method for identifying different critical points in TMD monolayers was discussed previously (23). This Σ state is also observed in tr-ARPES and visualized with an enhanced contrast in Fig. 4a albeit with a much lower count rate than the K point.

Shown in Fig. 4a are color rendering images of intensity versus energy and momentum for states at K and Σ respectively acquired at 300 K with a pump energy of 2.2 eV (above the quasi-particle gap) at two different delay times (0.13 ps and 1.27 ps). We integrate the photoemission signal over a range in the momentum space (pink and blue dashed boxes) to enhance the signal to noise. The signal strength at Σ is more than an order of magnitude smaller than that at K. We first analyze the dynamics as a function of energy. From the EDCs shown in Fig. 4b, one can observe that at the K point, the spectral peak shows a small downward shift by 34 ± 3 meV after t > 1 ps. On the other hand, a significantly larger downward shift of 180 ± 26 meV from 0.34 eV to 0.16 eV is observed at the Σ point. The 180 meV shift is likely an under-estimate since the EDC spectra are integrated over ± 0.2 $ps$. The initial energy location, labeled as $Σ_c$, is 0.16 ± 0.04 eV above the CBM, consistent with the energy differences measured using STS within the error bar. The excited state dynamics at Σ is examined in energy as a function of time delay in Fig. 4c, illustrating a rapid shift to a level (labeled as $Σ^*$) which is ~ 0.2 eV below the initial $Σ_c$ state within 0.5 ps. The $Σ^*$ state then decay with a slower time scale of ~4 ps. This large downward energy shift of ~0.2 eV for $Σ^*$ in reference to $Σ_c$ implies either the formation of a bound state with a binding energy of ~ 0.2 eV or a significant band renormalization occurring in sub-picosecond scale. This point is further discussed below along with the K-valley dynamics.

The excited state dynamics at the K point is quite distinct from those in the Σ valley. A small transient energy shift of ~34 meV is accompanied by clear changes in the dispersion. As shown in the inset of Fig. 5a, an upward dispersion is measured at t = 0.4 ps. We analyze the time-resolved dynamics acquired at four different momentum regions near the K point as shown in Fig. 5a. For region 1 (as labeled in the inset) with the highest energy and largest momentum deviation from K, the dynamics peaks at the shortest delay time of ~0.26 ps and exhibits the fast decay of 0.5 ps. As one approaches K, the dynamics systematically peak at later times and then decay with longer time constants (regions 2 and 3). Finally, at the K point the dynamics peaks at ~1 ps and decays with a time constant of ~10 ps. This trend is consistent with a hot electron decay down to the local minimum in k-space. In Fig. 5b we show two K-valley integrated EDC spectra acquired at t = 0.4 ps and 1.0 ps, respectively. The fitted areas below each spectral peak are nearly identical, within 2%, indicating that the all the hot electrons in the K-valley are cooled to the energy minimum at K and the additional contribution from intervalley-scattered $\Sigma_c$ electrons is minor.

The latter conclusion is further corroborated with the time-dependent dynamics acquired at $K_c$ and $\Sigma_c$ respectively in Fig. 5c. The time evolution of $\Sigma_c$ shows a rapid rise time of ~160 fs [the root mean square (RMS) width of the cross-correlation of pump-probe pulses is 140 fs], indicating rapid population of free carrier states at $\Sigma_c$ (*26*) which then shift within ~ 0.5 ps to $\Sigma^*$. On the other hand, the time evolution of $K_c$ shows a long rise time of ~0.4 ps and decays within ~10 ps. While inter-valley scattering from $\Sigma_c$ to $K_c$ may contribute to the hot carrier distribution in the K valley (*27, 28*), its contribution should be minimal given that the intensity at $\Sigma_c$ is smaller by an order of magnitude. This analysis shows that the K-valley dynamics is dominated by hot electron cooling with the above gap excitation.

On the other hand, the rapid evolution of $\Sigma_c$ to $\Sigma^*$ suggests a different origin. We note that in a recent study of tr-ARPES of WSe2, a similar rapid shift has been observed for both K- and Σ-valleys which was interpreted as the formation of exciton (K) and dark exciton (Σ) respectively (*13*). In our case, however, the large energy shift is observed only for Σ-valley and not for K-valley. One possibility is that bound states are formed at both valleys but the K-valley exciton signal is overshadowed by the free carrier signal at K-valley whose population is orders of magnitude higher. It is known that photoluminescence of monolayer $MoS_2$ on graphite is quenched by more than two orders of magnitude in comparison to $MoS_2$ monolayers on insulating substrates (*24*).

In conclusion, by successfully creating large scale monolayer $MoS_2$ with well-aligned in-plane orientation and atomically clean surface, we investigate its ground and excited state electronic structure. The equilibrium band structure is measured via ARPES and is confirmed to be consistent with DFT calculations. Using extreme-ultraviolet tr-ARPES as a tool to access the unoccupied conduction band across momentum space, we found a remarkable agreement on the electronic gap using STM and photoemission measurements. Moreover, tr-ARPES provides access to valley-specific excited-state dynamics. While a nearly instantaneous energy shift (~0.2 eV) to a lower energy state is observed at the $\Sigma$ point the dynamics at the K valley is dominated by hot carrier decay over a few picoseconds for the above-gap excitation. Future experiments that fully explore the complex electron dynamics in time, energy, and momentum will guide the search for new valley and exciton physics and devices based on TMD monolayers and heterostructures.

## Methods

### ARPES and trARPES

The regular ARPES measurements of the occupied band structure (Fig. 2) were performed at room temperature. A Helium lamp (21.2 and 40.8 eV) was used as the photon source. The spectra were collected using a Scienta R3000 analyzer. During the Helium lamp operation, the pressure was maintained under $6 \times 10^{-10}$ Torr. XUV-trARPES measurements were conducted at Berkeley Lab. The measurements for the unoccupied band structure were performed at 300 K and 80 K. 2.2-eV pump pulses and p-polarized 22.3-eV probe pulses with 25-kHz repetition rate were used in the measurements. Averaged pump power was measured to be 25 mW in front of the chamber entrance window. We estimated the maximum incident fluence on the sample to be 40 $\mu J/cm^2$, leading to an absorbed fluence of ~26 $\mu J/cm^2$ (excitation density ~7×10$^{11}$ cm$^{-2}$) after taking into account surface reflections. The pump-probe cross-correlation RMS width was estimated as 140 fs via the fastest photoemission dynamics. The setup highest energy resolution as measured from Au is 60.4 meV (*19*). During the XUV-trARPES measurements, the pressure was maintained at $3 \times 10^{-10}$ Torr.

### Density functional theory(DFT) calculation

The DFT calculation was carried out using the Quantum Espresso package (*29*, *30*) to obtain the free-standing monolayer MoS$_2$ band structure. Full relativistic projector augmented wave (PAW) pseudopotential was engaged to include spin-orbit coupling and plane wave basis with 40 Ry plane wave cutoff energy were employed. Perdew-Burke-Ernzerhof (PBE) form of exchange-correlation functional was used in the generalized gradient approximation (GGA). The Brillouin zone sampling was performed on a 12 × 12 × 1 k-point grid. The atomic coordinates and lattice constant were fully relaxed before calculating the electronic band structure.

**Acknowledgement**

This research was primarily supported by the National Science Foundation through the Center for Dynamics and Control of Materials: an NSF MRSEC under Cooperative Agreement No. DMR-1720595 and the facility supported by MRSEC. C.K.S. also acknowledge NSF-DMR 1808751, Welch Foundation F-1672 and US Airforce FA2386-18-1-4097. Work at Lawrence Berkeley National Laboratory by Y.L. and R.A.K. including tr-ARPES was supported by the U.S. Department of Energy (DOE), Office of Science, Office of Basic Energy Sciences, Materials Sciences and Engineering Division under Contract No. DE-AC02-05-CH11231 (Ultrafast Materials Science Program No. KC2203). L.W.-H.C. acknowledges the support from the Ministry of Science and Technology of Taiwan (MOST-107-2112-M-009-024-MY3 and MOST-108-2119-M-009-011-MY3) and the Center for Emergent Functional Matter Science (CEFMS) of NCTU supported by the Ministry of Education of Taiwan.

**Author contributions**

C.K.S. and W.H.C. originate the idea of creating this material platform. L.S.L. and W.C.C. prepared the samples. W.L. created a platform for UHV investigation. Regular ARPES measurements for characterization of the samples are performed by W.L. and STM measurements by M.L. tr-ARPES collaboration was initiated by R.A.K., X.L., and C.K.S. tr-ARPES and LEED measurements are carried out by W.L. and Y.L. The data analysis is carried out by W.L. with the supervision of R.A.K. and C.K.S. Paper is written by W.L. and C.K.S. with inputs from all other co-authors.

**Data availability**

The data that support the findings of this study are available from the corresponding author upon reasonable request.

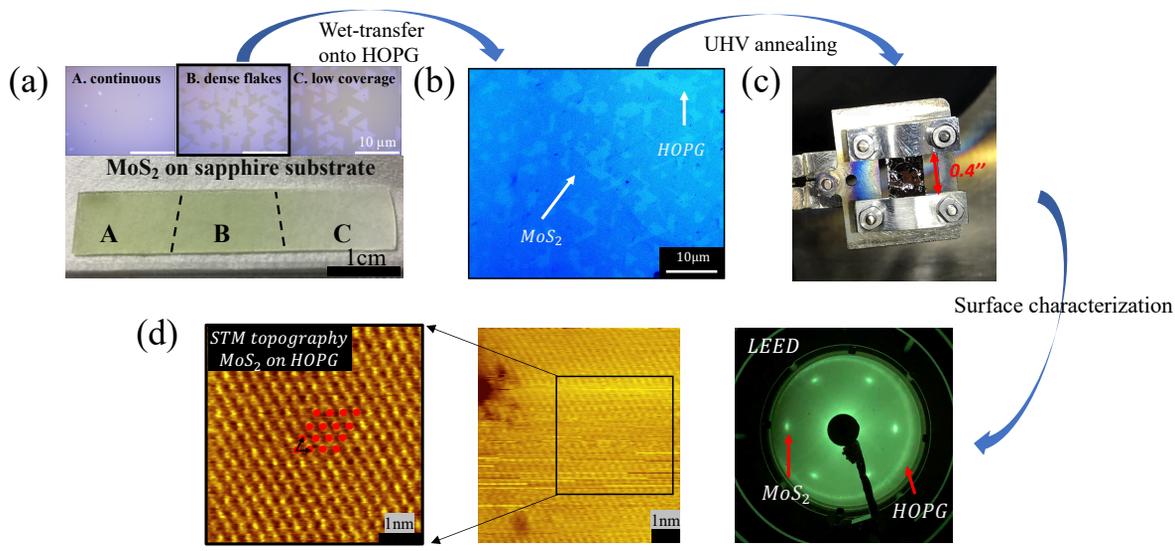

**Figure 1** Sample preparation and characterization. (a) Highly oriented MoS$_2$ monolayers are grown on a wafer scale sapphire substrate via substrate-guided CVD growth method. Zoomed-in optical images of the CVD-grown MoS$_2$ monolayers on the sapphire substrate in region A, B and C are in the upper panels. (b) Optical images after wet-transfer onto HOPG substrates. Due to the different reflectivity of the substrates in optical images, MoS$_2$ monolayers appear as bright regions on a sapphire substrate, but dark on HOPG substrate. (c) The lateral sample size (0.4 inch) is limited by the physical size of the sample holder. A clean sample surface following UHV annealing permits surface probes such as STM and ARPES. (d) The high quality of the sample on a HOPG substrate is confirmed with LEED at 68 eV and STM topography images.

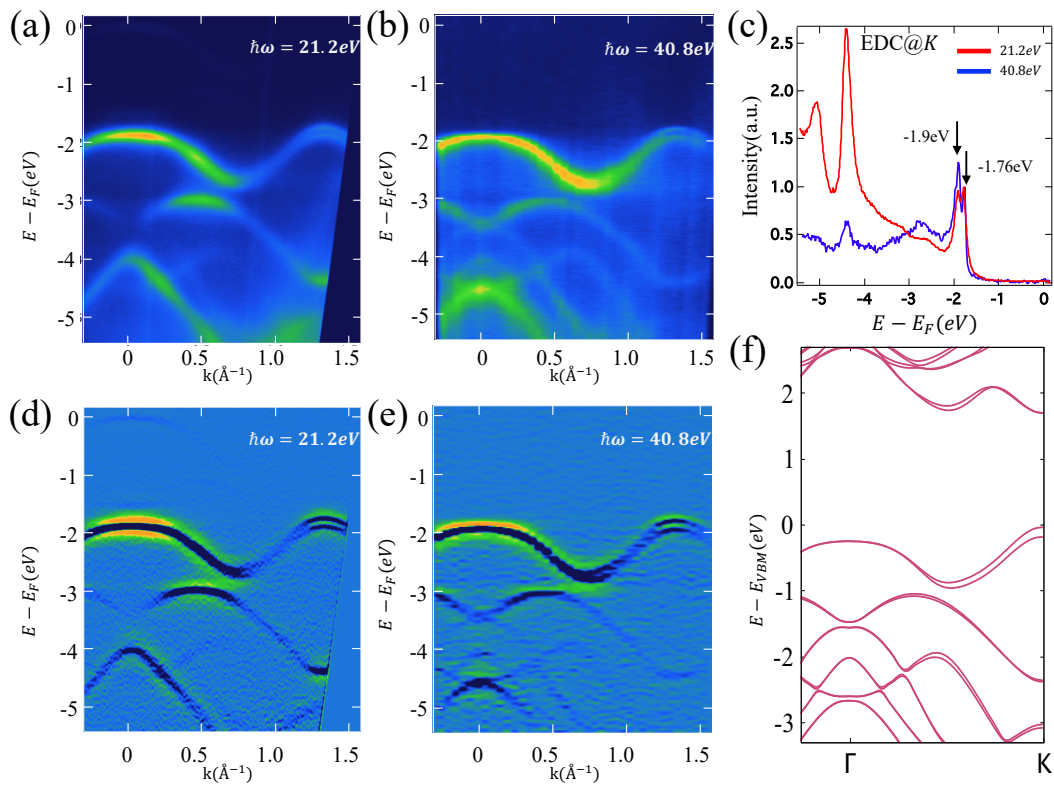

**Figure 2** Regular ARPES measurements on monolayer MoS$_2$ on HOPG. Different photon energies of a helium lamp are used to measure occupied band structure of MoS$_2$ monolayer, (a) $\hbar\omega = 21.2$ eV (b) $\hbar\omega = 40.8$ eV respectively. For a better visualization, second derivative images of (a) and (b) are represented in (d) and (e) respectively. (c) EDC at K point integrated over $\pm 0.025 \text{Å}^{-1}$. Red and blue solid lines indicate 21.2 and 40.8 eV photon energy, respectively. The spin orbit splitting was measured to be $140 \pm 4$ meV. (f) Electronic band structure of a free-standing MoS$_2$ monolayer obtained by DFT calculations using the Quantum espresso package.

Fig. 3

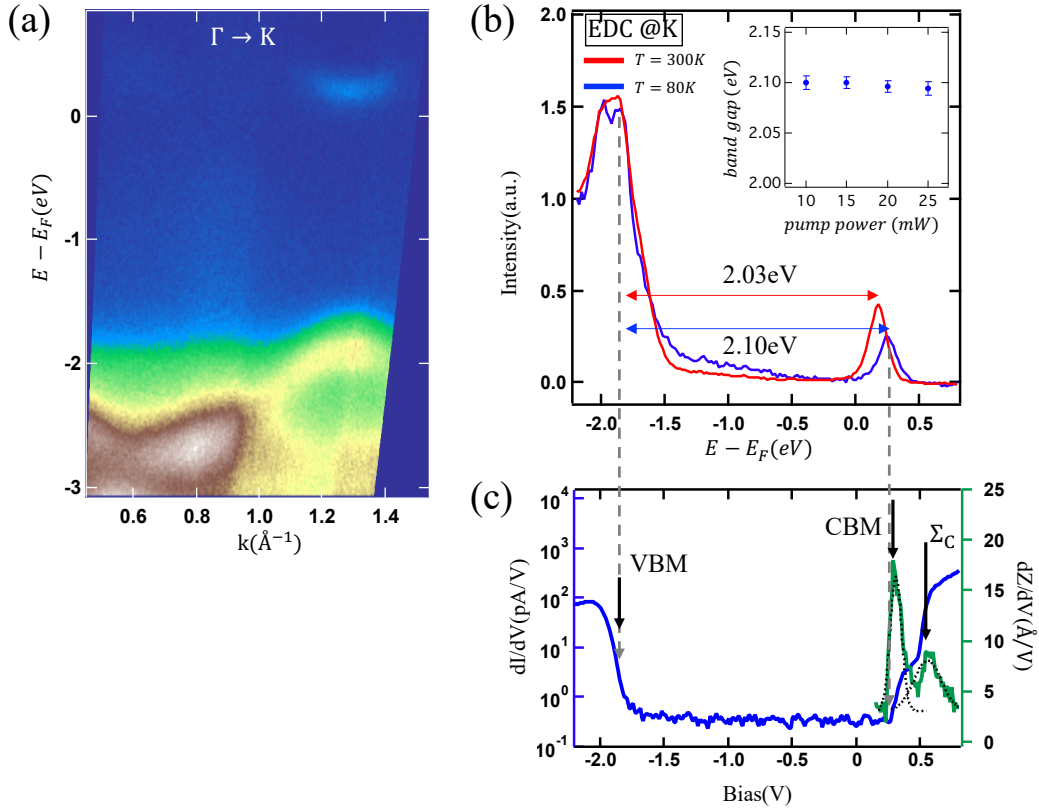

**Figure 3** XUV tr-ARPES measurements and scanning tunneling spectroscopy (STS) on monolayer $MoS_2$ on HOPG. (a) E-k dispersion along $\Gamma - K$ direction taken at 0.4ps and integrated over $\pm 0.2$ ps. Conduction band above the Fermi energy at K point is clearly observed. (b) EDC at K point, taken at different sample temperatures, at t = 1.27 ps and integrated over $\pm 0.5$ ps. The gap size is $2.03 \pm 0.01$ eV at 300 K and increases to $2.10 \pm 0.01$ eV at 80 K. Error bars represent standard deviation(SD) from Gaussian fitting. (c) STS dI/dV curve (blue) and dI/dZ curve(green) taken at liquid nitrogen temperature show a good agreement with tr-ARPES measurement regarding the VBM and CBM peak position.

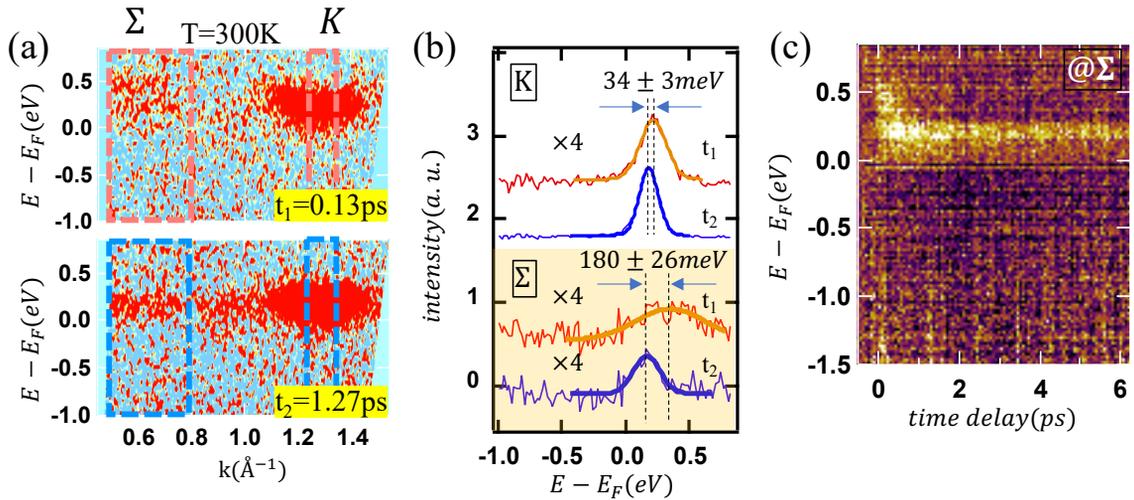

**Figure 4** Ultrafast carrier relaxations at Σ and K points measured at 300K. (a)E-k images are taken at different delay times with an integration time window of ± 0.2 ps to improve the signal-to-noise ratio. The images visualize a large shift at Σ point. (b) We extract EDCs from the dashed box regions in (a) to quantitatively show the energy shifts at Σ and K points. (c) Integrated electron density in dashed box ($\Delta k = 0.25$ Å$^{-1}$) at Σ point along the time delay. All data was subtracted from the equilibrium state to remove the background signal. A movie of the data is included in the Supplementary.

Fig. 5

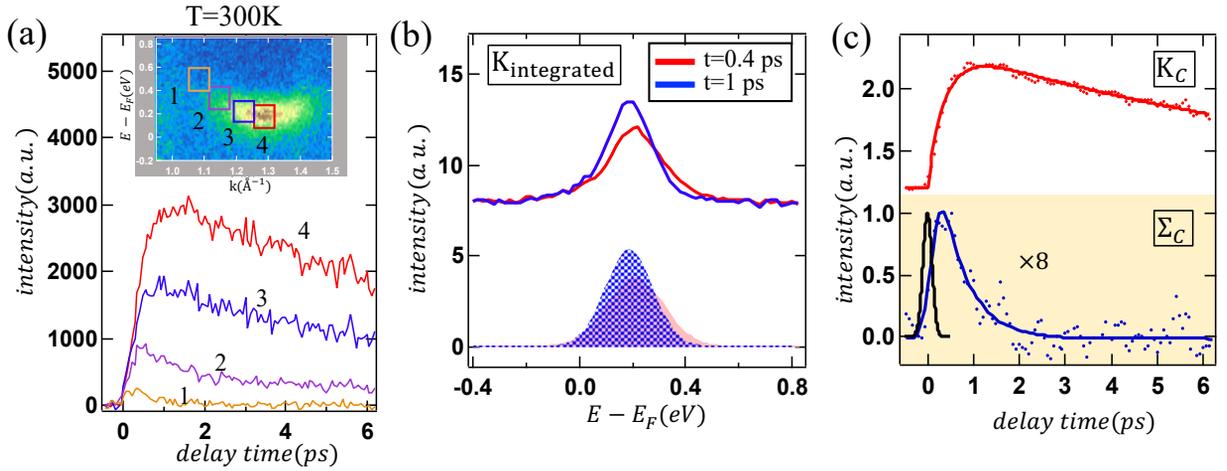

**Figure 5** Intra- and Inter- valley scattering at 300K. (a) Electron intensity versus time delay at different positions near the K point in the CB. The signals are integrated in colored boxes in the inset. (b) Momentum integrated EDC at K point. The area of the curve stands for total number of photoelectrons detected in the CB K-valley. (c) Time delayed electron density distribution integrated over $\Delta k = 0.25$ Å$^{-1}$ at Σ and K conduction bands. The black Gaussian function ($\sigma = 140$ fs) represents the cross-correlation between the pump and probe pulses. Time delayed signal from K point conduction band was fitted with a biexponential function (solid red curve) and Σ point conduction band was fitted with a single exponential decay function convoluted with a Gaussian function (solid blue curve).